\documentclass[12pt]{article}
\usepackage{graphicx}
\usepackage{epsfig}

\def\.{\cdot}
\def\la{\lambda}

\def\o{\over}

\def\a{\alpha}

\def\b{\beta}
\def\d{\delta}

\def\f{\phi}

\def\La{\Lambda}

\def\x{\xi}

\def\p{\partial}

\def\+{\bigoplus}
\def\D{\Delta}

\def\({\left(}
\def\){\right)}
\def\[{\left[}
\def\]{\right]}
\def\l.{\left.}
\def\r.{\right.}

\def\free{\mathcal{F}(J)}
\def\Tr{\mathrm{Tr}}
\def\dif{d^d\!}
\def\scl{{\scriptstyle(}}
\def\scr{{\scriptstyle)}}
\def\Ddot{\dot{\D}_t( x,y)}
\def\Del{\D_t( x,y)}
\def\Dinv{\D^{-1}_t( x,y)}
\def\sss{\scriptscriptstyle}
\def\pk{p^{\sss (i_k)}_{\sss k,\, j_k}}

\def\be{\begin{equation}}
\def\ee{\end{equation}}
\def\bea{\begin{eqnarray}&&}
\def\eea{\end{eqnarray}}
\def\nn{\nonumber \\ &&}
\def\nnn{\nonumber \\ }
\def\acca{\right.\nn\left.}
\def\ber{\begin{array}}
\def\eer{\end{array}}

\newcommand\grscale{0.15}

\begin{document}

\begin{titlepage}

\noindent
February, 2002\hfill

\vskip 1in
\begin{center}
\def\thefootnote{\fnsymbol{footnote}}
{\large \bf The Wilson-Polchinski Renormalization Group Equation in the Planar Limit}

\vskip 0.3in

C. Becchi \footnote{E-Mail: becchi@ge.infn.it}, S. Giusto \footnote{E-Mail: giusto@ge.infn.it} and C. Imbimbo 
\footnote{E-Mail: imbimbo@ge.infn.it} 

\vskip .2in

     {\em Dipartimento di Fisica, Universit\`a di Genova\\
          and\\ Istituto Nazionale di Fisica Nucleare, Sezione di Genova\\
              via Dodecaneso 33, I-16146, Genoa, Italy}

\end{center}

\vskip .4in

\begin{abstract}
We derive the Wilson-Polchinski RG equation in the planar
limit. We explain that the equation necessarily involves also non-planar amplitudes with sphere topology, which 
represent multi-trace contributions to the effective action. The resulting RG equation turns out to be of the 
Hamilton-Jacobi type since loop effects manifest themselves through terms which are linear in first order
derivatives of the effective action with respect to the sources. We briefly outline applications to
renormalization of non-commutative field theories, matrix models with external sources and holography.
\end{abstract}

\vfill

\end{titlepage}
\setcounter{footnote}{0}


\section{Introduction}
Planar field theories introduced years ago in the seminal work
of `t Hooft \cite{tHooft} have recently enjoyed renewed interest 
in connection with holography \cite{Maldacena},~\cite{Witten} and non-commutative field theory
\cite{Seiberg} (for a review and references see \cite{Douglas}).

The holographic conjecture relates 
$SU(N)$ (supersymmetric) gauge theory in $d$ dimensions
to string theory in $(d+1)$-dimensional anti de Sitter 
space. 
Consider a $(d+1)$-dimensional ball $B_{d+1}$  in the interior of AdS space,
whose boundary $S_d$ is a $d$-dimensional sphere homotopic to the
AdS boundary.  The action of AdS supergravity evaluated
on the solution of the classical field equations with boundary
values prescribed on $S_d$ and integrated on 
$B_{d+1}$ is a functional of the field boundary values: Maldacena's
conjecture identifies 
this functional with the generating functional of 
correlators of gauge-invariant composite
operators of the $SU(N)$ theory in the large $N$ limit and large `t Hooft
gauge coupling, renormalized at a scale which is related to the radial
distance of the sphere $S_d$ from the AdS boundary.

This intriguing correspondence between the AdS radial coordinate and
the renormalization scale of the gauge theory has suggested to some
authors \cite{Verlinde} the possibility that the Wilson-Polchinski
renormalization group equation of the $SU(N)$ quantum field theory in the 
large $N$ limit might correspond  to the classical equations of motion of the 
supergravity theory. The RG Wilson-Polchinski equation \cite{Wilson},~\cite{Polchinski}
is a differential 
equation for the effective action of first order in the renormalization scale:
therefore, in the scenario proposed in \cite{Verlinde}, it is 
put in relation with the supergravity field equations written
in the Hamilton-Jacobi form, with the AdS radial direction identified
with the time coordinate. Since radial slices are not space-like surfaces 
in AdS, this identification appears problematic. 
At any rate, any attempt to make this proposal more concrete
requires the knowledge of the large $N$ limit
of the RG Wilson-Polchinski equation. This is what we do in this article.

The RG Wilson-Polchinski equation in the planar limit might also
be of interest in the study of the renormalization properties of the
non-commutative quantum field theories which emerge in a certain limit
of open string theory. It has in fact been observed~\cite{Seiberg}
that the perturbative expansion for these theories reduces, in the limit of 
large non-commutative parameter $|\Theta|$, to planar diagrams. It is therefore 
relevant to understand if a RG equation involving only planar
diagrams can be written for such theories. 

\begin{figure}
\centerline{
\begin{picture}(250,100)(0,0)
\put(0,22){\line(0,1){46}}\put(3,17){\line(1,0){94}}
\put(3,71){\line(1,0){94}}\put(101,22){\line(0,1){46}}
\put(6,24){\framebox(40,40){}} \put(54,24){\framebox(40,40){}}
\put(50,0){\makebox(0,0){\footnotesize (a)}}\put(190,0){\makebox(0,0){
\footnotesize (b)}} 
\put(140,22){\line(0,1){49}}\put(143,17){\line(1,0){94}}
\put(140,71){\line(1,0){100}}\put(240,22){\line(0,1){49}}
\put(146,24){\framebox(40,40){}}
\put(194,24){\line(1,0){40}}\put(194,24){\line(0,1){16}}
\put(194,48){\line(0,1){16}}
\put(194,64){\line(1,0){40}}\put(234,24){\line(0,1){16}}
\put(234,48){\line(0,1){16}}
\end{picture}}
\caption[x] {\footnotesize Planar (a) and non-planar (b) diagrams 
in non\--com\-muta\-tive theory }
\label{f:planar}
\end{figure}
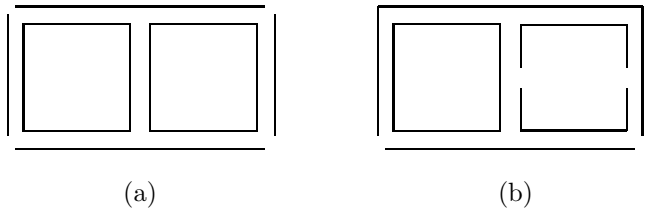

Let us comment on the meaning of planar diagrams in
the context of field theories discussed in~\cite{Seiberg},
whose non-commutative structure is described by the Moyal product. 
Feynman diagrams of non-commutative theories have two 
equivalent descriptions: in one description propagators are single
lines and lines coming out of vertices are ordered.
In the other description propagators are double lines carrying opposite orientations. 
In both pictures, a Feynman diagram uniquely 
determines a Riemann surface on which it can be drawn: this surface
is oriented, due to the orientation of propagator lines, and it is closed, i.e. without
boundaries, as long as one restricts oneself to theories without single lines (quarks). 
We will always refer to the double line picture in the following. In this picture diagrams 
corresponding to
amputated and connected
Green functions are called planar if they can be drawn
on the sphere {\it and} if all the external legs are
attached to the same line. Figure~\ref{f:planar} shows two examples
of planar $(a)$  and non-planar $(b)$ diagrams in non-commutative theory
with cubic interaction, both of which can be drawn on a sphere: 
in this figure external legs are represented by breaks
in the propagator lines. With this definition, 
non-planar diagrams like $(b)$ are 
exponentially suppressed in Moyal non-commutative
theories in the limit of large non-commutativity (see Appendix A). 
The effective action appearing in the 
RG group
equation is the generating functional of amputated connected Green functions.
It is relatively easy 
to convince oneself and it will be explained in detail in
Section 1 that one cannot write
a RG group equation involving exclusively planar amplitudes: 
diagrams like $(b)$ must be included in the ``planar''
Wilson-Polchinski equation.

In $SU(N)$ matrix theories sphere diagrams like $(a)$ correspond
to terms in the Wilsonian effective action which are single
traces of the matrix source while diagrams like $(b)$ are associated
to multi-traces.  
The large $N$ limit for the Wilsonian action is to be taken by
keeping fixed {\it both} the `t Hooft couplings {\it and} 
the normalized trace invariants of the external matrix source. 
In this limit sphere diagrams --- both of $(a)$ and $(b)$
type --- are of the same
(leading) order in the $1/N$ expansion. This makes clear
why the planar limit of RG equation that we derive 
describes all sphere diagrams, irrespectively of where the external
legs are attached.

This paper is organized as follows: in Section 2 a diagrammatic
derivation of the planar RG equation (\ref{planarRG}) for 0-dimensional
non-commutative theory is presented. It is interesting that this equation,
unlike the usual RG Wilson-Polchinski equation, is of the Hamilton-Jacobi type.
The first derivative with respect to the renormalization scale of the
Wilsonian action equals a sum of two terms:  the first is the classical
one and is quadratic in the field derivatives of the effective action.
The second term, which encodes the quantum corrections, 
is {\it linear} in field derivatives of the action: this is
in contrast with the usual RG equation for which the quantum
term involves second order field derivatives of the action. 
In Section 3 the same equation is derived by functional methods,
by taking the large $N$ limit of the Wilson-Polchinski equation
for scalar $N\times N$ matrix models.
In Section 4 a formulation of the equation in terms of 
eigenvalues of the matrix source is also given: in these variables
the planar RG equation is the Hamilton-Jacobi equation of a system of
$N$ {\it free} particles in 1 dimension. This reformulation leads
to a compact expression for the effective action in the large $N$ limit. 
In Section 5 we write down the planar RG equation for a
scalar matrix theory in $d$ dimensions. Finally in Section 6 we state our
conclusions. 

In the work of Ref.~\cite{MiaoLi} a large $N$ differential equation 
for the generating functional of the correlators of single-trace composite operators
of a zero-dimensional matrix model is derived. This generating functional might be thought of as 
the zero-dimensional analogue of the on-shell supergravity action which enters the holographic 
conjecture in the formulation of \cite{Witten}. In Appendix B we rederive this equation
in a language which is closer to the one of the present paper and at the same time we
clear up a few imprecisions present in the equation of \cite{MiaoLi}. It is important however 
to emphasize that this equation is not an equation for the Wilson-Polchinski effective action
and, if generalized to higher dimensions, it does not encode the renormalization properties
of the corresponding quantum field theory.  

\section{The ``Planar'' RG Equation from Feynman Diagrams} 

Let us consider a $d$-dimensional scalar field theory, whose field $\phi(x)$
takes values in some non-commutative space. We can think
of this space either as the space of $N\times N$ hermitian matrices
or as the non-commutative algebra associated with the Moyal
product. Let 
$h^{\scriptscriptstyle planar}_{t,t_0}(p_{1},\ldots,p_{n})$ be the 
planar contribution
to the $n$-point connected and amputated Green function
computed with a regulated propagator $\Delta_{t,t_0}(x,y)$ which depends
both on the ultra-violet scale $t_0\equiv \Lambda_0^{-2}$ and the running (infra-red)
scale $t\equiv \Lambda^{-2}$. As we explained in the introduction 
$h^{\scriptscriptstyle planar}_{t,t_0}(p_{1},\ldots,p_{n})$
receives contributions from diagrams of type $(a)$ of Figure~\ref{f:planar}.

The Wilson-Polchinski RG equation describes the dependence on the
infra-red scale $t$ of
the generating functional $H_t[\varphi]$ of 
(both planar and non-planar) regulated, amputated and 
connected Green functions. Since the dependence of the Green functions on
the ultra-violet cut-off $t_0$ never enters the discussion, henceforth
we will drop explicit reference to $t_0$. One might hope that a
consistent truncation of the Wilson-Polchinski equation 
to planar Green functions
would lead to an evolution equation for the planar
functional generator $H^{\scriptscriptstyle planar}_t[\varphi]$
\be 
H^{\scriptscriptstyle planar}_t [\f]\equiv\sum_{n=1}^{\infty}{1\o n}\int\!\prod_{i=1}^{n}d^dp_{i}\
  \d \Bigl(\sum_{j=1}^{n}p_{i}\Bigr)\,h^{\scriptscriptstyle planar}_t(p_{1},\ldots,p_{n})
\langle \varphi(p_{1})\ldots\varphi(p_{n})\rangle 
\label{1}
\ee 
where $\langle \cdot \rangle$  is the trace on the non-commutative
algebra. We will now explain why such a truncation does not exist:
in order to write down a ``planar '' RG equation one must extend
the functional in Eq. (\ref{1}) to include multi-trace terms.

Let us recall the diagrammatic interpretation of the ordinary (commutative)
RG equation\footnote{Notice that, in order to avoid alternating signs, we have adopted a 
sign convention which is opposite to the standard one.}. The $t$-derivative of $H_t$
is the formal sum of all the diagrams obtained from those
contributing to $H_t$ by replacing one of the internal propagators
$\D_t$ with its derivative $\dot\D_t$, i.e. by
inserting $\dot\D_t$ between the new external
legs created after cutting the propagator line. This leads to the
diagrammatic equation displayed in Figure~\ref{f:ordinaryRG}.
In this picture a box with a label $\a$ on the lower right-hand side
corner represents an amputated connected Green function with $\a$ 
external legs. A crossed line denotes the insertion of the $\dot\D_t$ factor. 
The two diagrams on the r.h.s. of the equation correspond
to the two possible cases: in the first case the removed
propagator line lies on a tree line of the skeleton graph while in the second
case the propagator belongs to a loop.

\begin{figure}[h]
\centerline{
\begin{picture}(250,50)(0,0)
\put(0,10){\makebox(0,0){$\p_t$}}
\put(7,3){\framebox(40,18){}}
\put(51,5){\makebox(0,0){\tiny$\a$}}
\put(70,12){\makebox(0,0){$\quad ={1\o 2}\!\sum_{\scriptscriptstyle \b}\quad\,$}}
\put(161,5){\makebox(0,0){\tiny $\a-\b+2$}}
\put(88,3){\framebox(18,18){}}
\put(126,3){\framebox(18,18){}}
\put(110,5){\makebox(0,0){\tiny $\b$}}
\put(106,11){\line(1,0){20}}
\put(116,11){\makebox(0,0){$\scriptscriptstyle \rm X$}}
\put(190,12){\makebox(0,0){$\!\!\!\!\! +\,\,{1\o 2}$}}
\put(200,3){\framebox(40,18){}}
\put(200,22){\line(2,1){20}}
\put(220,32){\line(2,-1){20}}
\put(220,31){\makebox(0,0){$ \scriptscriptstyle \rm X$}}
\put(243,5){\makebox(0,0){\tiny $\,\a$}}
\end{picture}}
\caption[x]{\footnotesize The ordinary RG evolution equation}
\label{f:ordinaryRG}
\end{figure}

Considering now the case of non-commutative or matrix theories, planar diagrams in the double-line
picture have a preferred line, that will be called {\it peripheral}.
This is the line to which all the external legs are attached: external
legs of planar diagrams have therefore a natural cyclic order.
Consider now the second term on the r.h.s. of the diagrammatic
equation in Figure~\ref{f:ordinaryRG}: when the external legs of a
planar amplitude which are joined by the $\dot{\Delta}_t$ are not
contiguous, this term induces a non-vanishing contribution to the
$t$-derivative of non-planar amplitudes. These are special non-planar
amplitudes, like $(b)$ of Figure~\ref{f:planar}: they have sphere
topology but have external legs attached to more than one single
line. Let us name {\it holes} the lines, other than the peripheral
line, to which external legs are attached.  It is clear now that if we
want to write a RG equation for the planar graphs we must, for
consistency, include also all the sphere graphs with any number of
holes to which an arbitrary number of external lines are attached.  We
might call these graphs {\it swiss-cheese} (SC) for their resemblance
to swiss-cheese slices when drawn on a plane.

\begin{figure}[b]
\centerline{
\begin{picture}(100,60)(0,-10)
\put(0,0){\framebox(80,40){}}
\put(86,5){\makebox(0,0){$\scriptstyle \a$}}
\put(40,-10){\makebox(0,0){\footnotesize (a)}} 
\end{picture}
\begin{picture}(100,60)(-25,-10)
\put(50,15){\circle{20}{$\scriptstyle \a_1$}}
\put(0,0){\framebox(80,40){}}
\put(87,5){\makebox(0,0){$\scriptstyle \a_0$}}
\put(40,-10){\makebox(0,0){\footnotesize (b)}}
\end{picture}}
\caption[x]{\footnotesize SC diagrams with 0 (a) and 1 (b) holes}
\label{f:examples}
\end{figure}
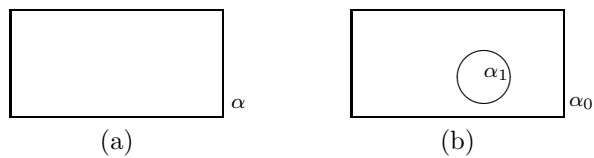

Let us picture an $\a$-point planar, connected and amputated amplitude with a box,
whose perimeter represents the peripheral line labelled
by an index $\a$ on the right-hand side lower corner. SC amplitudes
with $n$ holes, $\a_0$ external legs attached to the peripheral
line and $\a_i$ external legs attached to the
$i$-th hole (with $i=1,\ldots,n$) are represented by a box
with $n$ internal circular holes: the index $\a_0$ is associated with the perimeter 
of the box while $\a_i$'s label the internal circles. 
Two examples are depicted in Figure~\ref{f:examples}. Note that we call amplitude
the sum of all diagrams with the same structure of external legs. Therefore
the interior of boxes in Figure~\ref{f:examples} should be thought as filled with all
possible nets of propagators. Propagators will be now double lines and insertions of $\dot\D_t$ 
will be indicated by crosses: 
\begin{picture}(20,10)(0,0)
\put(0,3){\line(1,0){16}}
\put(0,1){\line(1,0){16}}
\put(8,2.5){\makebox(0,0){$\scriptscriptstyle \rm X$}}
\end{picture}. Let us also distinguish two kinds of propagators --- external
and internal.  External propagators have one line belonging to the
peripheral line, internal propagators are all the others. 

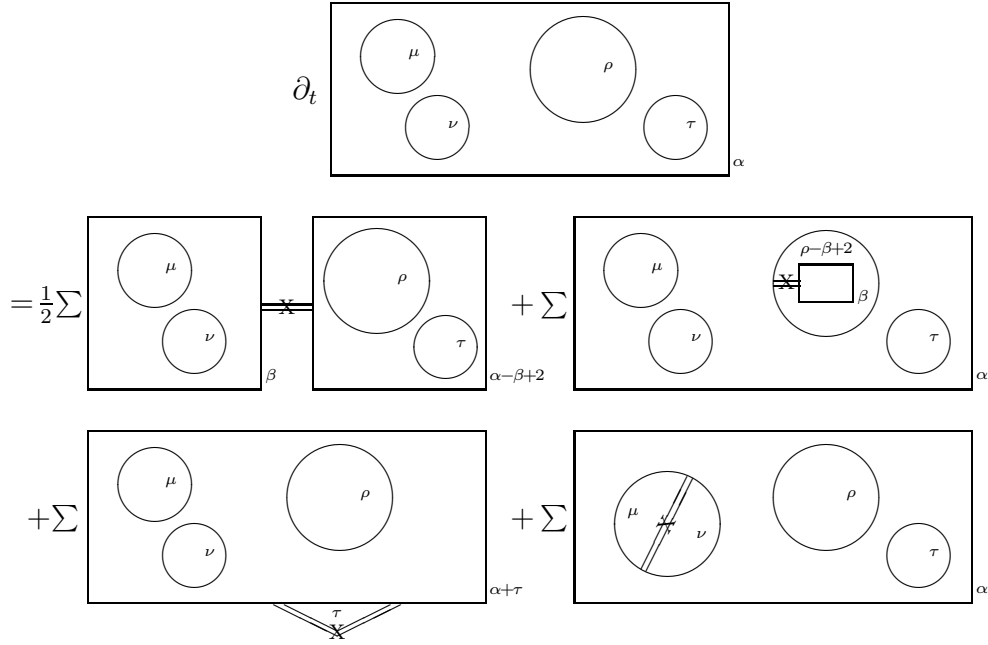
\begin{figure}
\begin{center}
\begin{picture}(180,80)(0,0)
\put(0,32){\makebox(0,0){$\p_t$}}
\put(50,18){\circle{24}{\tiny $\,\,\,\nu$}}
\put(35,45){\circle{26}{\tiny $\,\,\,\mu$}}
\put(105,40){\circle{50}{\tiny $\,\,\,\,\,\,\rho$}}
\put(140,18){\circle{24}{\tiny $\,\,\,\tau$}}
\put(10,0){\framebox(150,65){}}
\put(164,5){\makebox(0,0){\tiny $\a$}}
\end{picture}

\begin{picture}(180,80)(0,0)
\put(0,32){\makebox(-12,0){$=\!{1\o 2}\!\sum$}}
\put(50,18){\circle{24}{\tiny $\,\,\,\nu$}}
\put(35,45){\circle{26}{\tiny $\,\,\,\mu$}}
\put(172,5){\makebox(0,0){\tiny $\a\!\!-\!\!\b\!\!+\!\!2$}}
\put(119,41){\circle{50}{\tiny $\,\,\,\,\,\,\rho$}}
\put(145,16){\circle{24}{\tiny $\,\,\,\tau$}}
\put(10,0){\framebox(65,65){}}
\put(95,0){\framebox(65,65){}}
\put(79,5){\makebox(0,0){\tiny $\b$}}
\put(75,32){\line(1,0){20}}
\put(75,30){\line(1,0){20}}
\put(85,31){\makebox(0,0){$\scriptstyle \rm X$}}
\end{picture}
\begin{picture}(180,80)(0,0)
\put(0,32){\makebox(-7,0){$+\sum$}}
\put(10,0){\framebox(150,65){}}
\put(50,18){\circle{24}{\tiny $\,\,\,\nu$}}
\put(35,45){\circle{26}{\tiny $\,\,\,\mu$}}
\put(140,18){\circle{24}{\tiny $\,\,\,\tau$}}
\put(105,40){\circle{50}}
\put(164,5){\makebox(0,0){\tiny $\a$}}
\put(85,41){\line(1,0){10}}
\put(85,39){\line(1,0){10}}
\put(90,40.5){\makebox(0,0){$\scriptstyle \rm X$}}
\put(95,33){\framebox(20,14){}}
\put(119,35){\makebox(0,0){\tiny $\b$}}
\put(105,53){\makebox(0,0){\tiny $\rho\!\!-\!\!\b\!\!+\!\!2$}}
\end{picture}

\begin{picture}(180,80)(0,0)
\put(0,32){\makebox(-7,0){$+\!\sum$}}
\put(50,18){\circle{24}{\tiny $\,\,\,\nu$}}
\put(35,45){\circle{26}{\tiny $\,\,\,\mu$}}
\put(168,5){\makebox(0,0){\tiny $\a\!\!+\!\!\tau$}}
\put(105,40){\circle{50}{\tiny $\,\,\,\,\,\,\rho$}}
\put(80,-.5){\line(2,-1){24}}\put(84,-.5){\line(2,-1){20}}
\put(104,-10.5){\line(2,1){20}}\put(104,-12.5){\line(2,1){24}}
\put(10,0){\framebox(150,65){}}
\put(104,-11){\makebox(0,0){$\scriptstyle \rm X$}}
\put(104,-4){\makebox(0,0){\tiny $\tau$}}
\end{picture}
\begin{picture}(180,80)(0,0)
\put(0,32){\makebox(-7,0){$+\sum$}}
\put(45,30){\circle{40}}
\put(35,13){\line(1,2){17.5}}
\put(37,12){\line(1,2){17.5}}
\put(32,34){\makebox(0,0){\tiny $\mu$}}
\put(58,26){\makebox(0,0){\tiny $\nu$}}
\put(44.5,30){\makebox(0,0){\rotatebox{60}{$\scriptstyle \rm X$}}}
\put(164,5){\makebox(0,0){\tiny $\a$}}
\put(105,40){\circle{50}{\tiny $\,\,\,\,\,\,\rho$}}
\put(140,18){\circle{24}{\tiny $\,\,\,\tau$}}
\put(10,0){\framebox(150,65){}}
\end{picture}
\end{center}
\caption[x]{The ``planar'' RG equation}
\label{f:scrg}
\end{figure}

Given these graphical rules, the diagrammatic RG equation for SC
amplitudes which replaces Figure~\ref{f:ordinaryRG} is displayed in
Figure~\ref{f:scrg}. The four terms on the r.h.s.  of this equation
correspond to the four possible positions of the removed propagator
inside the graph. In the first and the second term the cut propagator
connects two otherwise disconnected graphs; in the third and fourth
term removing the propagator leaves the diagram connected.  The
removed propagator of terms one and three is external, that of terms
two and four is internal.

Terms one and two on the r.h.s. of the equation in Figure~\ref{f:scrg}
replace the first (classical) term on the r.h.s. of the ordinary RG
equation in Figure~\ref{f:ordinaryRG}.  Terms three and four
correspond to the loop term of ordinary RG flow (the last term of
Figure~\ref{f:ordinaryRG}). It is the third term on the
r.h.s. of the RG equation in Figure~\ref{f:scrg} that generates
non-planar diagrams starting from planar ones.

We now want to translate the graphical ``planar'' RG equation in Figure~\ref{f:scrg}
into analytic form. This requires the introduction of a functional generator
for SC amplitudes generalizing the one for planar amplitudes given
in Eq. (\ref{1}). As Eq. (\ref{1}) shows, planar amplitudes 
are represented in the effective action by 
a single trace of the product of the sources. 
Likewise, each hole of a SC amplitude is associated with 
the trace of the product of as many sources as legs attached to the hole.
Hence, SC non-planar amplitudes are associated with multi-traces, 
one trace for each hole and one trace for the
peripheral line. It is then clear that, when writing
down the effective action,  there is no distinction
between holes and peripheral line:
we can characterize the SC amplitude by the number $m_k$ of
traces of products of $k$ sources without the need to distinguish 
the trace associated with the peripheral line from the traces
associated with the holes.  Of course diagrammatically this corresponds to the fact
that SC graphs can be drawn on a sphere where the peripheral line is just another hole.
Henceforth we will refer to SC amplitudes when drawn on a sphere as multi-trace amplitudes. 
More explicitely, if we call $f_t(n,\,\{m_k\})$ the SC amplitude with $n$ peripheral legs
and $m_k$ $k$-leg holes and $h_t(\{ m_k \})$ the multi-trace amplitude with $m_k$ $k$-leg holes ---
we are suppressing explicit reference to external momenta for notational simplicity --- then one has
\be
{1\over n}\,f_t(n,\,\{m_k\}) = (m_n+1)\,h_t(\{ m_k + \d_{k,n}\})
\ee

\begin{figure}
\begin{center}
\raisebox{13pt}{$\p_t$}
\includegraphics*[scale=\grscale, clip=false]{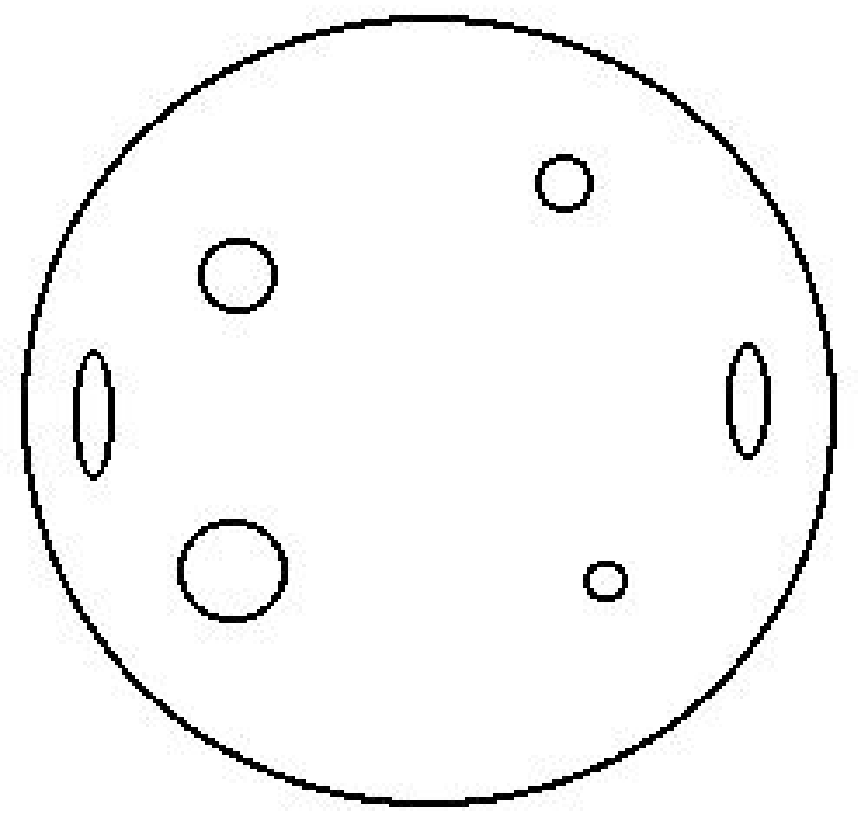}
\raisebox{13pt}{$= {1\over 2}\sum\ $}
\includegraphics*[scale=\grscale, clip=false]{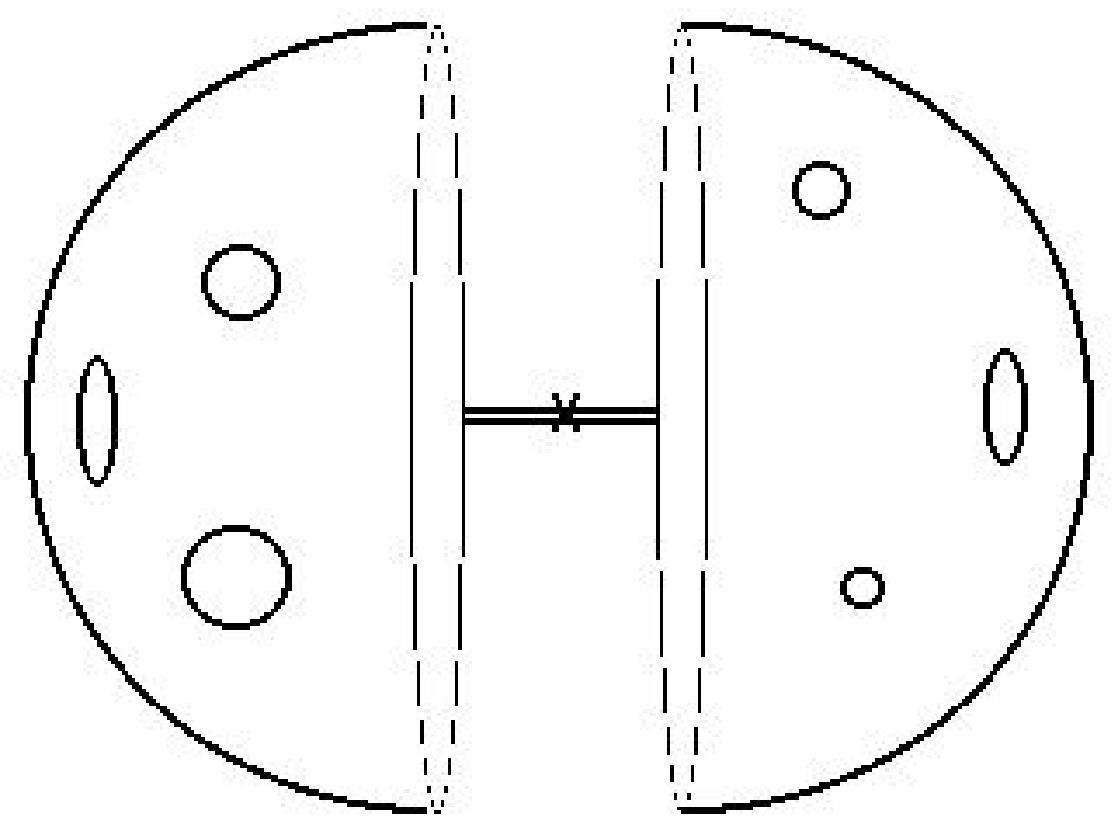}
\raisebox{13pt}{$+\,\sum\ $}\includegraphics*[scale=\grscale, 
clip=false]{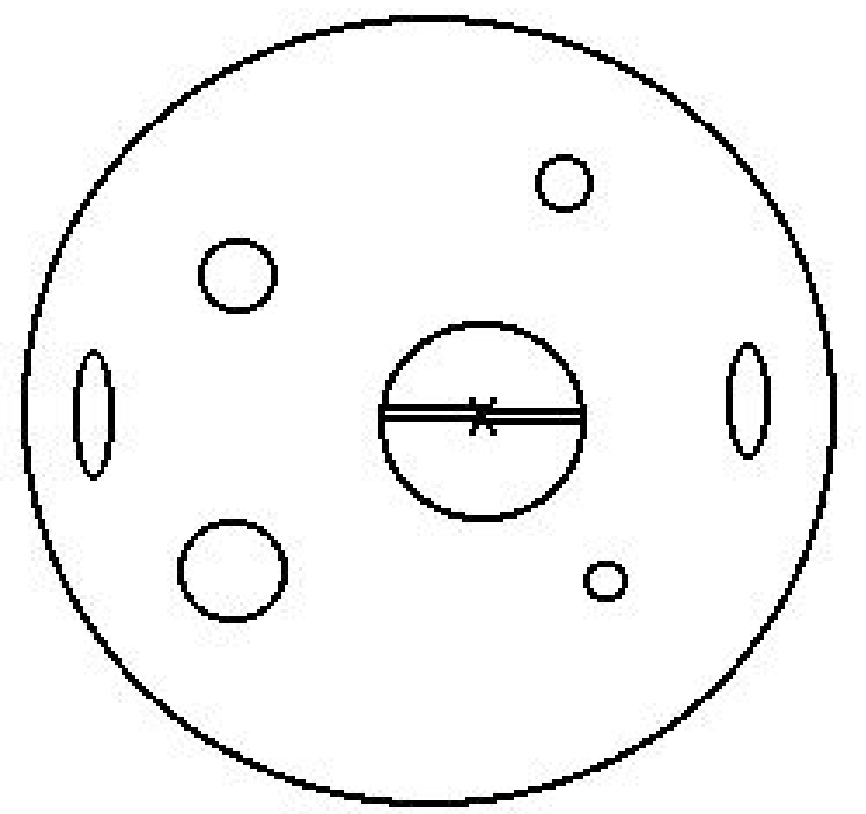}
\caption[x]{\footnotesize The ``planar'' RG equation on the sphere}
\label{f:Nrg}
\end{center}
\end{figure}

If we represent multi-trace amplitudes with spheres with holes the RG equation in
Figure~\ref{f:scrg} becomes the graphical equation of Figure~\ref{f:Nrg}.
Since the distinction between internal and external propagators is immaterial
when we draw the diagram on the sphere, the first and the last pair 
of terms on the r.h.s. of the equation in Figure~\ref{f:scrg} become respectively
the first and the second term in the equation of Figure~\ref{f:Nrg}.
When the theory is a $N\times N$ matrix theory,
the sphere representation of the RG equation in  Figure~\ref{f:Nrg}  
is perhaps more natural, since in the $1/N$ expansion sphere diagrams with holes are
precisely the leading ($O(N^2)$) contribution to the effective Wilsonian
action. In the next section where we will derive the ``planar'' RG equation by
functional methods we will adopt the matrix point of view.

Let $\pk$ be the $j_k$-th momentum (with $j_k=1,\ldots,k$)
which flows into the $i_k$ hole (with $i_k=1,\ldots,m_k$) of a 
multi-trace graph corresponding to the connected amputed amplitude 
$h^{\{m_k\}}_{t} (\pk)$.
The Wilsonian effective action writes
\be
H_t[\varphi]\!\equiv
\!\!\sum_{\{m_k\}}\!\int\!\!\!\prod_{\sss k,i_k,j_k}
\!\!\! dp^{\sss (i_k)}_{{\scriptscriptstyle k,\, j_k}}\
  \d \bigl(\!\!\sum_{\sss k,i_k,j_k}\pk\bigr)\,
h^{\{m_k\}}_{t} (\pk)\prod_k\prod_{i_k=1}^{m_k}
\langle\varphi (p^{\sss (i_k)}_{\sss k,\, 1})\cdots
\varphi (p^{\sss (i_k)}_{\sss k,\, k })
\rangle
\label{scfunctional}
\ee

To simplify our task and notation we shall disregard momentarily 
the momenta, therefore limiting our analysis
to zero-dimensional non-commutative field theory. Obviously in this
case the propagator is simply a constant which we will identify with
the RG scale $t$ and ${\dot \Delta}_t$ reduces to $1$.  
The contribution of a given diagram to the corresponding
amplitude is just the product of the combinatorial factor of the diagram
with the required powers of the couplings and of the propagator.
It should be clear however that the structure of the RG equation  --- which is captured
by the graphical Equations Figures~\ref{f:scrg}, \ref{f:Nrg} ---
is independent of the dimension of space-time.  In Section 5 we will explicitly
write down the $d$-dimensional version of the RG equation. 

In $0$ dimensions the functional generator in Eq. \ref{scfunctional}
becomes
\be
H_t(\{y_{k}\})\equiv
\sum_{\{m_k\}} h_t(\{m_k\})\prod_{k=1}^{\infty}y_{k}^{m_{k}}
\label{hgen}
\ee
where 
\be
y_k \equiv \langle \varphi^{k}\rangle
\ee

The first term on the r.h.s of the diagrammatic equation in Figure~\ref{f:Nrg}
corresponds to the following contribution to $\partial_t H_t(\{y_k\})$
\bea
{1\o2} \sum_{k,\,k^\prime = 1}^{\infty}\sum_{\{m_p\}}\sum_{\{m^\prime_p\}}
k\,m_k\,k^\prime\,m^\prime_{k^\prime}\, h_t(\{m_p\})\,h_t(\{m^\prime_p\})\,
{y_{k+k^\prime-2}\over y_k\,y_{k^\prime}}
\prod_{p=1}^{\infty}y_{p}^{m_p+m^\prime_p}\nn 
= {1\o2}\sum_{k,\,k^\prime = 1}^{\infty} k\,k^\prime\,y_{k+k^\prime-2}\,
\partial_{y_k}H_t\,\partial_{y_{k^\prime}}H_t
\label{firstterm}
\eea
The summand on the l.h.s. of the equation above corresponds to  
the joining of two multi-trace diagrams labelled respectively by the sequences 
$\{m_p\}$ and $\{m^\prime_p\}$; the crossed propagator
connecting the two diagrams goes from a hole of the first
diagram with $k$ external legs to a hole of the second diagram
with $k^\prime$ legs. The factor $k\,m_k\,k^\prime\,m^\prime_{k^\prime}$ 
counts the different ways of joining two such diagrams. The resulting 
connected multi-trace diagram has a hole with $k+k^\prime-2$ legs 
replacing the two holes with $k$ and $k^\prime$ legs: 
this is accounted for by the ${y_{k+k^\prime-2}\over y_k\,y_{k^\prime}}$ 
factor. The overall ${1\over 2}$ factor is due to the fact
that exchanging the two multi-trace diagrams produces the same connected
amplitude. 

The second term on the r.h.s of the diagrammatic equation 
in Figure~\ref{f:Nrg} yields:
\bea
{1\o2} \sum_{k = 1}^{\infty}\sum_{i=0}^{k-2}\sum_{\{m_p\}}
k\,m_k\, h_t(\{m_p\})\,
{y_i\,y_{k-2-i}\over y_{k}}
\prod_{p=1}^{\infty}y_{p}^{m_p}\nn 
= {1\o2}\sum_{k = 1}^{\infty}\sum_{i=0}^{k-2} 
k\, y_i\,y_{k-2-i}\,\partial_{y_{k}}H_t
\label{secondterm}
\eea
The summand in the l.h.s. of Eq. (\ref{secondterm})
describes the process of joining
two legs of the same hole of a multi-trace diagram labelled
by the sequence $\{m_p\}$. Let $k$ be the number
of external legs of this hole. 
The possible resulting multi-trace diagrams have one less hole with $k$ legs
and two more holes with $i$ and $k-2-i$ legs respectively,
where $i=0,\ldots, k-2$. This explains the ${y_i\,y_{k-2-i}\over y_{k}}$
factor. Note that when $i=0$ and $i=k-2$  one of the two holes 
has no external legs attached to it: if we define $y_0=1$, the
factor  ${y_i\,y_{k-2-i}\over y_{k}}$ is the correct one 
for these cases too. 
Given $k$, there are $k\,m_k$ ways to choose one of the
two legs to be joined; the inequivalent choices for the second leg 
are labelled by the integer $i$ with $i=0, \ldots, \bigl[{k-2\over 2}\bigr]$.
Since the factor ${y_i\,y_{k-2-i}\over y_{k}}$ is invariant under
the substitution $i \to k-2-i$,
we can equivalently sum over $i$ in the range $[0,\, k-2]$
and divide by ${1\over 2}$. 

In conclusion the RG equation for the functional generator of
multi-trace amplitudes in $0$ dimensions reads:
\be
\partial_t H_t =
{1\o2}\sum_{k,\,k^\prime = 1}^{\infty} k\,k^\prime\,y_{k+k^\prime-2}\,
\partial_{y_k}H_t\,\partial_{y_{k^\prime}}H_t\,
+\, {1\o2}\sum_{k = 1}^{\infty}\sum_{i=0}^{k-2} 
k\, y_i\,y_{k-2-i}\,\partial_{y_{k}}H_t
\label{planarRG}
\ee

The striking feature of this RG equation is that only first derivatives
with respect to the sources $y_k$ appear: thus the equation is
of the Hamilton-Jacobi type. This is to be contrasted
with the Wilson-Polchinski equation for ordinary (commutative)
field theory which contains second order derivatives with respect
to the sources. For example, the ordinary RG equation of Figure~\ref{f:ordinaryRG}
for a scalar field in $0$ dimensions reads
\be
\partial_t H_t = {1\over 2} \bigl(\partial_\varphi H_t\bigr)^2 + {1\over 2}
\partial_\varphi^2 H_t
\label{analiticordinaryRG}
\ee
where $H_t(\varphi) = \sum_{n=0}^\infty {1\over n!}\, h_t(n)\,\varphi^n$ 
is the generating functional of connected amputated
amplitudes $h_t(n)$ for the scalar field whose source is $\varphi$.
The first ``classical'' term in the r.h.s. of the ordinary
RG equation is of the same form of its analog in the non-commutative
RG: the difference between the two equations lies in the ``loop''
term, which becomes first order in source derivatives 
when going to the non-commutative context.

A solution of the RG equation (\ref{planarRG}) is specified by
giving the initial condition $H_t(\{y_k\})\vert_{t=0}$. This should
be interpreted as the ``bare'' or classical interaction Lagrangian\footnote{Due to our
sign convention, the initial condition on $H_t$ is actually minus the interaction
Lagrangian.},
which is usually specified as a {\it linear} function of the
$y_k$'s. Note however that even if the initial condition is linear in
the $y_k$, $H_t$ for $t\not=0$ will necessary be non-linear:
this reflects the impossibility of writing down the RG equation
for the planar functional generator (\ref{1}) that we have emphasized
at the beginning at this section.

To consider a concrete example, let us choose the initial condition
corresponding to a $\lambda\langle\phi^4\rangle$ theory:
\be
H_t(\{y_k\})\vert_{t=0}= {1\over 4}\lambda\, y_4 
\label{initial}
\ee
Let us remark that when the intial condition is even under  ``parity''
transformation $y_k \to (-1)^k y_k$ we can consistently truncate 
our equation to an effective action $H^{\scriptscriptstyle (+)}_t \equiv 
H_t\vert_{\scriptscriptstyle y_{2k+1}=0}$ depending only on the even $y_{2k}$ variables. 
This is due to the circumstance that the RG equation both conserves parity and
is first order in $y_k$-derivatives. The truncated RG equation is

\be
\partial_t H^{\scriptscriptstyle (+)}_t =
2\!\!\!\sum_{k,\,k^\prime = 1}^{\infty}\!\! 
k\,k^\prime\,y_{ 2k+2k^\prime-2}\,
\partial_{y_{2k}}H^{\scriptscriptstyle (+)}_t\,\partial_{y_{2k^\prime}}
H^{\scriptscriptstyle (+)}_t + \sum_{k = 1}^{\infty}\!\sum_{i=0}^{2k-2} 
\!k\, y_{2i}\,y_{2k-2-2i}\,\partial_{y_{2k}}H^{\scriptscriptstyle (+)}_t
\label{evenplanarRG}
\ee 

It is straightforward to compute the solution of Eq. (\ref{evenplanarRG}) 
with initial condition (\ref{initial}) as perturbative expansion in $t$:
 
\bea 
H_t^{\scriptscriptstyle (+)}(\lambda; \{y_{k}\}) = 
{1\o4}\lambda\, y_4 +\( \lambda\, y_2 + {1\o2}\lambda^2
y_6 \)t +\({3\o2}\lambda^3 y_8+{5\o2}\lambda^2 y_4 \acca +{3\o4}\lambda^2
y^2_2+{1\o2}\lambda\)t^2 +\({11\o2}\lambda^4 y_{10}+{28\o3}\lambda^3
y_6+{9\o2}\lambda^2 y_2 + 5\,\lambda^3 y_2\, y_4\) t^3 \nn +
\({91\o4}\lambda^5 y_{12}+{165\o4}\lambda^4 y_8+{45\o2}\lambda^3 y_4+21\,\lambda^4
y_2\, y_6 + {27\o2}\lambda^3 y^2_2 \acca + {75\o8}\lambda^4 y^2_4+{9\o8}\lambda^2\)t^4
+ \(27\,\lambda^3 y_2 + {27\over 2}\lambda^4 y_2^3 + 135\,\lambda^4 y_2\, y_4
+ 126\,\lambda^4 y_6 \acca + 84\,\lambda^5\, y_4\, y_6 + 99\,\lambda^5 y_2\, y_8 + 
{1001\over 5}\lambda^5 y_{10} + 102\,\lambda^6 y_{14}\) t^5 +O(t^6)
\label{hsol}
\eea

The interested reader can verify that the coefficients of the single 
terms in this sum count the number of the corresponding  Feynman diagrams multiplied by their
combinatorial factor. Owing to the choice of the factor ${1\o4}$ in 
the interaction Eq.~(\ref{initial}) the combinatorial factor of the diagram 
is the order of its rotation symmetry group in the plane: for example the planar 
version of the diagram
\begin{picture}(20,10)(0,0)\put(0,2){\line(1,0){20}}
\put(10,2){\circle{6}}\end{picture} has combinatorial factor ${1\o2}$.

The effective
action of the $\lambda \phi^4$ theory is invariant under the following rescaling
\be
t\to \sigma\, t \qquad  \lambda \to \sigma^{-2} \lambda 
\qquad y_{2k}\to \sigma^{k} y_{2k} 
\label{rescaling}
\ee
where $\sigma$ is a real parameter. This reflects the invariance
of the classical action under transformation of  $t$ and $\lambda$ as in
(\ref{rescaling}) and simultaneous rescaling of the classical field given
by: $\phi \to \sqrt{\sigma}\, \phi$. Hence $H_t^{\scriptscriptstyle (+)}$
is a function of the variables $\tau \equiv t^2 \lambda$ and 
$\eta_{2k} \equiv t^{-k} y_{2k}$. 
The equation for ${\cal H}(\tau; \{ \eta_{2k}\})\equiv 
H_t^{\scriptscriptstyle (+)}(\lambda; \{y_{2k}\})$ is
\bea
2\tau \partial_\tau {\cal H} = 2\!\!\!\sum_{k,\,k^\prime = 1}^{\infty}\!\! 
k\,k^\prime\,\eta_{2k+2k^\prime-2}\,
\partial_{\eta_{2k}}{\cal H}\,\partial_{\eta_{2k^\prime}}
{\cal H}\nn 
\qquad \qquad + \sum_{k = 1}^{\infty}\!\sum_{i=0}^{2k-2} 
\!k\, \eta_{2i}\,\eta_{2k-2-2i}\,\partial_{\eta_{2k}}{\cal H}
+\sum_{k=1}^\infty k\, \eta_{2k}\, \partial_{\eta_{2k}}{\cal H}
\label{strongplanarRG}
\eea 
This form of the equation is a convenient starting point for studying the strong
coupling limit of the theory, $\tau \to \infty$. In this limit
the interaction term ${1\over 4}\tau\, \phi^4$ in the classical
action dominates, so that it is natural to rescale the field as 
$\phi\to \tau^{-{1\over 4}}\,\phi$ and to work with the sources
$\zeta_{2k} \equiv \tau^{-{k\over 2}}\eta_{2k}$.  The leading
term in the effective action equals
$-{1\over 4}\log \tau - {1\over 2}\tau^{1\over 2}\zeta_2$,
as it can been easily understood by considering, for example,
the path integral representation of the Wilson-Polchinski action. 
Hence ${\cal H}$ admits the following
asymptotic expansion for $\tau \to \infty$
\bea
{\cal H}(\tau; \{\eta_{2k}\}) = -{1\over 4} \log \tau - {1\over 2}
\tau^{1\over 2}\zeta_2 + \widetilde{\cal H}(\tau; \{\zeta_{2k}\})\nn
\qquad \qquad\quad\,\,  = - {1\over 4} \log \tau - {1\over 2}
\tau^{1\over 2}\zeta_2 + \tilde{h}_0(\{\zeta_{2k}\}) +
O(\tau^{-{1\over 2}})
\label{asymptotic}
\eea
The function $\tilde{h}_0(\{\zeta_{2k}\})$ is to be interpreted as the
initial condition of the strong coupling limit and has to be computed by 
independent methods. Once $\tilde{h}_0(\{\zeta_{2k}\})$ is specified, the
equation for $\widetilde{\cal H}$
\be
2\tau \partial_\tau \widetilde{\cal H} = \tau^{-{1\over 2}}
\Bigl[2\!\!\!\sum_{k,\,k^\prime = 1}^{\infty}\!\! 
k\,k^\prime\,\zeta_{2k+2k^\prime-2}\,
\partial_{\zeta_{2k}}\widetilde{\cal H}\,\partial_{\zeta_{2k^\prime}}
\widetilde{\cal H}
+ \sum_{k = 1}^{\infty}\!\sum_{i=0}^{2k-2} 
\!k\, \zeta_{2i}\,\zeta_{2k-2-2i}\,\partial_{\zeta_{2k}}
\widetilde{\cal H}\Bigr]
\label{superstrongplanarRG}
\ee
determines recursively the subleading $O(\tau^{-{1\over 2}})$ terms.

\section{Functional Derivation of Large N Wilson-Pol\-chinski RG Equation}

The generating functional $\free$ of the connected Green functions
of the $N\times N$ matrix scalar theory is

\be
{\rm e}^{-N^2 \free } \equiv \mathcal{N} \int d\Phi\, {\rm e}^{-N \bigl[S(\Phi) -\Tr\, J \Phi \bigr]}    
\label{freeenergy}
\ee
where $J$ is the $N\times N$ matrix source for the quantum matrix field $\Phi$
and $S(\Phi)$ is the classical action
\be
S(\Phi) = {1\over 2t} \Tr\, \Phi^2 - \Tr\, V(\Phi)
\ee
For example we can take, as in the previous section,
$V(\Phi) = {\lambda\over 4}\Phi^4$.  The normalization factor
$\mathcal{N}$ is 
\be 
\mathcal{N}^{-1} = \int d\Phi\, {\rm e}^{-{N\over 2t} \Tr\, \Phi^2}=
\Bigl({ \pi t\over N}\Bigr)^{N^2/2}
\ee

The Wilsonian effective action $H_t(\varphi)$ is the 
generating functional of the connected {\em amputated} Green functions.
It is related to $\free$ as follows:
\be
\free = - H_t(\varphi) - {1\over 2 N t} \Tr\, \varphi^2 
\ee
where $J = {1\over t}\varphi$.

The Wilson-Polchinski RG equation determines the dependence of $H_t(\varphi)$
on the ``propagator'' $t$. The large $N$ limit should be taken
keeping the invariants 
\be
y_{k} \equiv {1\over N}\Tr\, \varphi^{k}
\label{invariants}
\ee
where $k=1,2,\ldots$, {\it finite} as $N\to \infty$. 

Taking the derivative with respect to $t$ of both sides of Eq.
(\ref{freeenergy}) one obtains
\be
\partial_t H_t(\varphi) - {1\over 2t^2 N} \Tr\, \varphi^2 = 
-{1\over 2t} 
+{1\over 2 N t^2}\langle\Tr\, \Phi^2 \rangle  - {1\over N t^2}\langle
\Tr\, \varphi\, \Phi \rangle 
\ee 
where 
\be
\langle O(\Phi)\rangle = {\int\! d\Phi\, 
{\rm e}^{-N[S(\Phi)-{1\over t}\Tr\,\varphi\, \Phi]}
\, O(\Phi) \over \int\! d\Phi\, {\rm e}^{-N[S(\Phi)-{1\over t}\Tr\,
\varphi\, \Phi]}}
\ee
Using
\bea
{1\over N}\langle \Tr\, \varphi\,\Phi\rangle = - t\, \Tr\,\varphi\, 
{\partial\mathcal{F}\over \partial \varphi}\nn
{1\over N}\langle \Tr\, \Phi^2 \rangle= -{t^2\over N}\, \Tr\, 
{\partial^2\mathcal{F}\over \partial \varphi^2} + t^2 N\,\Tr\, 
\Bigl({\partial \mathcal{F}\over \partial \varphi}\Bigr)^2 
\eea
we get
\be
\partial_t H_t = {N\over 2}\,\Tr\, 
\Bigl({\partial H_t\over \partial \varphi}\Bigr)^2 
+ {1\over 2 N}\, \Tr\, 
{\partial^2 H_t \over \partial \varphi^2}  
\label{finiteNRG}
\ee 
which is the finite $N$ RG equation. In the large N limit we should
keep fixed the variables $y_k$. Hence, let us express the derivatives
with respect to $\varphi$ in terms of $y_k$-derivatives:
\be
{\partial H_t\over \partial \varphi} = \sum_k  k {\varphi^{k-1}\over N}
\partial_{y_k}\, H_t
\ee
and
\begin{eqnarray}
&&{1\over  N}\, \Tr\, 
{\partial^2 H_t(\varphi)\over \partial \varphi^2} = 
{1\over N^2} \sum_{k,\, k^\prime} k\, k^\prime y_{k+k^\prime-2}\,\partial_{y_k}
\partial_{y_{k^\prime}} H_t \nnn
&&\qquad\qquad\qquad\qquad + \sum_k  {k\over N^2} 
\Tr\, \Bigl({\partial \varphi^{k-1}\over\partial \varphi}\Bigr)
\,\partial_{y_k} H_t
\end{eqnarray}
In the large $N$ limit the first term in the r.h.s. of the equation
above goes to zero. The second term, instead, survives. Indeed,  
since
\be
\Tr\, \Bigl({\partial \varphi^{k-1}\over\partial \varphi}\Bigr)
= \sum_{i=0}^{k-2}\Tr\, \varphi^i\, \Tr\, \varphi^{k-2-i}
\ee
one has
\be
\sum_k  {k\over N^2} 
\Tr\, \Bigl({\partial \varphi^{k-1}\over\partial \varphi}\Bigr)
\,\partial_{y_k} H_t = \sum_k \sum_{i=0}^{k-2} k\, y_{i}\, y_{k-2-i}
\,\partial_{y_k} H_t
\ee
which is finite as $N\to\infty$.
In conclusion the RG equation in the large $N$ limit becomes
\be
\partial_t H_t = {1\over 2}\sum_{k,k^\prime} k\, k^\prime y_{k+k^\prime-2}
\,\partial_{y_k} H_t\, \partial_{y_{k^\prime}} H_t + {1\over 2}
\sum_k \sum_{i=0}^{k-2} k\, y_{i}\, y_{k-2-i}\,\partial_{y_k} H_t
\label{largeNRG}
\ee
in agreement with Eq. (\ref{planarRG}).

\section{Application to Matrix Models}

In the application of Eq. (\ref{largeNRG}) to $0$-dimensional matrix
models it is usually convenient to replace the $y_k$ variables with the
eigenvalues $\x_i$ of the matrix source $\varphi$. Starting from
\be
y_k ={1\over N} \sum_i \x^k_i
\ee
it easily shown that
\be
\Tr\, \Bigl({\partial H_t\over \partial \varphi}\Bigr)^2 = 
\sum_i \bigl(\partial_{\x_i}H_t \bigr)^2
\label{eigen1}
\ee
and that
\be
\Tr\, {\partial^2 H_t \over \partial \varphi^2} = 
\sum_i \partial_{\x_i}^2 H_t + \sum_{\scriptstyle i,\,j:\, i \not= j} {\partial_{\x_i} H_t
-\partial_{\x_j} H_t \over \x_i -\x_j}
\label{eigen2}
\ee
Going back to the finite $N$ RG equation (\ref{finiteNRG}), it rewrites therefore as follows:
\be
\partial_t H_t = {N\over 2}\,\sum_i \bigl(\partial_{\x_i}H_t \bigr)^2
+ {1\over 2 N}\,\sum_i \partial_{\x_i}^2 H_t +{1\over 2 N}\, \sum_{\scriptstyle i,\,j:\, i\not= j} 
{\partial_{\x_i} H_t
-\partial_{\x_j} H_t \over \x_i -\x_j}
\label{xifiniteNRG}
\ee 
One can re-absorb the term linear in $\x_i$'s derivatives by
defining
\be
{\hat H}_t = H_t + {1\over 4 N^2}\sum_{\scriptstyle i,\,j:\, i \not= j}\log(\x_i -\x_j)^2
\label{shiftedaction}
\ee
Thus
\be
\partial_t {\hat H}_t = 
{N\over 2}\,\sum_i \bigl(\partial_{\x_i}{\hat H}_t \bigr)^2
+ {1\over 2 N}\,\sum_i \partial_{\x_i}^2 {\hat H}_t 
\label{xifiniteshiftedNRG}
\ee 
In order to understand the large $N$ behaviour of the equation above, we introduce 
the eigenvalue density \cite{Parisi}
\be
\rho(\x)\equiv {1\over N} \sum_i \delta(\x -\x_i),
\ee
and rewrite Eq. (\ref{xifiniteshiftedNRG}) as follows
\be
\partial_t {\hat H}_t = 
{1\over 2}\,\int d\x\,  \rho(\x)\Biggl[\Bigl(\partial_{\x}{\delta {\hat H}_t\over
\delta\,\rho(\x)} \Bigr)^2
+ {1\over N^2}\,\partial_{\x}^2{\delta^2 H_t
\over\delta\,\rho(\x)^2}\Biggr]
\label{rhofiniteshiftedNRG}
\ee 
This makes clear that the second
term in the r.h.s. which is second order in field derivatives 
is {\it subleading} in the large
$N$-limit. 

Thus the planar limit of the $0$-dimensional RG equation
reduces just to the {\it free} Hamilton-Jacobi equation for a system of 
$N$ particles (with $N\to\infty$)
\be
\partial_t {\hat H}_t = {N\over 2}\,\sum_i \bigl(\partial_{\x_i}{\hat H}_t \bigr)^2
\ee

We can therefore express ${\hat H}_t$ in terms of the initial condition
\be
{\hat V}(\{\x_i\}) = \sum_{i} V(\x_i) + {1\over 4 N}\,\sum_{\scriptstyle i,\,j:\, i \not= j}\log(\x_i -\x_j)^2
\label{hatinitial}
\ee  
by means of the general integral of the free Hamilton-Jacobi equation:
\be
{\hat H}_t(\{\x_i\}) = {1\over N}\,{\hat V}(\{\zeta_i\}) - {t\over 2 N}\, \sum_{i}
\bigl(\partial_{\zeta_i}\,{\hat V}(\{\zeta_i\})\bigr)^2
\label{hateffectiveaction}
\ee
where the functions $\zeta_i(\x_i,t)$ are implicitely defined by the following algebraic
equation
\bea
\zeta_i = \x_i + t\,\partial_{\zeta_i}\,{\hat V}(\{\zeta_i\})\nn
\quad\, = \x_i + t\, V^\prime(\zeta_i) + {t\over N}\, \sum_{\scriptstyle j:\,j \not= i} 
{1\over \zeta_i -\zeta_j}
\label{algebraiceq}
\eea 
In conclusion, we have reduced the solution of a generic matrix model with interaction $V(\Phi)$ coupled
to an external source $\varphi$, in the large $N$ limit, to the solution of the algebraic equation 
(\ref{algebraiceq}). In terms of the solution $\zeta_i(\x_i,t)$ of this equation, 
the large $N$ effective action $H_t$ of the matrix model is
\bea
\label{finaleffectiveaction}
\!\!\!\!\!\!\!\!\!\!H_t = {1\over N}\,\sum_i V(\zeta_i) - {t\over 2 N}\,\sum_i (V^\prime(\zeta_i))^2
-{t\over 2 N^2}\!\sum_{\scriptstyle i,\,j:\,i \not= j}\!
{V^\prime(\zeta_i)-V^\prime(\zeta_j)\over \zeta_i -\zeta_j}\\ \nonumber
&& - {1\over 4 N^2}\! \sum_{\scriptstyle i,\,j:\,i \not= j}\!\log
\Biggl(1-t {V^\prime(\zeta_i)-V^\prime(\zeta_j)\over \zeta_i -\zeta_j}+ {t\over N}\!
\sum_{\scriptstyle k:\,k \not= i,\,j} {1\over (\zeta_i-\zeta_k)(\zeta_j-\zeta_k)}
\Biggr)^{\!\!2}
\eea
In deriving Eq. (\ref{finaleffectiveaction}) from Eqs. (\ref{shiftedaction}),
(\ref{hateffectiveaction}) and (\ref{hatinitial}) we have suppressed terms
\be
-{t\over 2 N^3}\,\sum_{{i,\,j,\,k\atop j,\,k \not = i}}\,{1\over \zeta_i - \zeta_j}\,{1\over \zeta_i - \zeta_k} = -{t\over 2 N^3}\,\sum_{\scriptstyle i,\,j:\,i \not= j}{1\over (\zeta_i - \zeta_j)^2}
\ee
and
\be
-{2t\over N}\,{1\over (\zeta_i-\zeta_j)^2}
\ee
because sub-leading in the $1/N$ expansion. Note, however, that terms of order $1/N^p$ with $p>0$ are 
still contained in equation (\ref{finaleffectiveaction}). 
 

\section{Extension to $d$ Space-Time Dimensions}

The generating functional $\mathcal{F}[J]$ of the regularized 
connected Green functions of the $d$-dimensional $N\times N$ matrix scalar 
theory is

\be
{\rm e}^{-N^2 \mathcal{F}[J] } \equiv \mathcal{N} \int [d\Phi]\, {\rm e}^{-N \bigl[S[\Phi] -\int\! \dif x\,\Tr\, J\scl x \scr  \Phi\scl x\scr  \bigr]}    
\label{freeenergyd}
\ee
where $J(x)$ is the $N\times N$ matrix source for the quantum matrix field 
$\Phi(x)$ and $S[\Phi]$ is the regularized classical action
\be
S[\Phi] = {1\over 2} \int\!\dif x\,\dif y\, \Dinv\, \Tr\, \Phi(x) \Phi(y)  - \int\!\dif x\, V(\Phi(x) )
\ee
As we explained in the previous section, $\Del$ is the propagator
regulated both in the IR and in the UV. We omit explicit reference
to the UV scale $t_0$ and denote by $t$ the infra-red cut-off:
 $\Del$ vanishes rapidly in the UV limit 
$t\to t_0$. The normalization factor $\mathcal{N}$ is 
\bea
\mathcal{N}^{-1} = \int [d\Phi]\, {\rm e}^{-{N\over 2}\! \int\!\dif x\,\dif y\,
\Dinv\,\Tr\, \Phi\scl x\scr \Phi\scl y\scr }\nn\qquad\,\,=
\Bigl({ \pi \mathrm{det}_{x,y}(\Del)\over N}\Bigr)^{N^2/2}
\eea

The Wilsonian effective action $H_t[\varphi]$ 
is related to $\mathcal{F}[J]$ as follows:
\be
\mathcal{F}[J] = -H_t[\varphi] - {1\over 2 N}\int\!\dif x\,\dif y\, 
\Dinv\, \Tr\, \varphi\scl x\scr \varphi\scl y\scr  
\ee
where $J(x) = \int\!\dif y \, \Dinv\varphi\scl y\scr $.

The Wilson-Polchinski RG equation determines the dependence of $H_t[\varphi]$
on the infra-red scale $t$. The large $N$ limit should be taken
by keeping the invariants 
\be
Y_{k}( x_1,\ldots,x_{k})  \equiv {1\over N}\Tr\, \varphi\scl x_1\scr \cdots\varphi
\scl x_{k}\scr 
\label{invariantsd}
\ee
where  $k = 1, 2,\ldots$, {\it finite} as $N\to \infty$. 
Going through the same steps as in the previous section, we obtain the 
finite $N$ RG equation
\be
\partial_t H_t \!=\!{N\over 2}\!\! \int\!\dif x\, \dif y\, \Ddot\Bigl[
\Tr\, {\d H_t\over \d \varphi\scl x\scr }{\d H_t\over \d \varphi\scl y\scr } 
+{1\over N^2}\, \Tr\, 
{\d^2 H_t\over \d \varphi\scl x\scr  \d \varphi\scl y\scr }\Bigr]  
\label{finitend}
\ee 
To take the large $N$ limit we must express the derivatives
with respect to $\varphi$ in terms of $Y_k$-derivatives:
\be
{\d H_t\over \d \varphi\scl x\scr } \!=\! \sum_k  {k\over N} \!\!\int\!\!d x_1\ldots d x_{k-1}\,
\varphi\scl x_1\scr \cdots\varphi\scl x_{k-1}\scr \,
{\d H_t \over \d Y_k( x, x_1,\ldots,x_{k-1}) }
\ee
For the same reasons that have been explained in the previous section, 
the second derivative term in Eq. (\ref{finitend}) 
reduces, in the large $N$ limit, to
\bea
\!\!\!\!\!\!{1\over N}\Tr\, 
{\d^2 H_t\over \d \varphi\scl x\scr\,  \d \varphi\scl y\scr } \to 
\sum_k \sum_{i=1}^{k-1} k \int\!\! d x_1\ldots d x_{k-1}\,\d\scl x_i-x\scr \nn
\times\,Y_{i-1}( x_1,\ldots, x_{i-1}) \,Y_{k-1-i}( x_{i+1},\ldots,x_{k-1}) \,
{\d H_t \over \d Y_k( y, x_1,\ldots,x_{k-1})  }\nonumber
\eea

In conclusion the RG equation in the large $N$ limit becomes
\bea
\partial_t H_t \! =\! {1\over 2} \!\int\!dx\,dy\, \Ddot \Biggl[\,
\sum_{k,k^\prime} k k^\prime \!\!\int \!\! 
d x_1\ldots d x_{k-1}\, d y_1\ldots d y_{k^\prime-1}\nn 
\qquad\qquad \times \, Y_{k+k^\prime-2}( x_1,\ldots,x_{k-1},y_1,\ldots,y_{k^\prime-1}) \nn 
\qquad\qquad \times \, {\d H_t\over \d Y_k( x,x_1,\ldots,x_{k-1}) }\, 
{\d H_t \over \d Y_{k^\prime}(y,y_1,\ldots,y_{k^\prime -1}) }+\nn 
\qquad\qquad + 
\sum_k \sum_{i=1}^{k-1} k \int\!\! d x_1\ldots d x_{k-1}
\,\d\scl x_i-x\scr \nn
\qquad\qquad \times \, Y_{i-1}( x_1,\ldots, x_{i-1}) \,Y_{k-1-i}( x_{i+1},\ldots,x_{k-1})\nn
\qquad\qquad \times \, {\d H_t \over \d Y_k( y, x_1,\ldots,x_{k-1})  }\,\Biggr]
\label{largenrgd}
\eea
This is still a Hamilton-Jacobi type equation for the system of multi-local sources 
$Y_k(x_1,\ldots,x_k)$. Due to the non-trivial $t$-dependence of $\Ddot$, this system
is described by a time-dependent Hamiltonian, in contrast to the zero-dimensional case.
Note that, even if we derived Eq. (\ref{largenrgd}) in the framework of $N \times N$ matrix theory,
the equation is valid for the non-commutative Moyal case as well.
 
\section{Conclusions}
In this article we have derived, both from diagrammatic and functional methods, the
``planar'' Wilson-Polchinski RG equation for a $d$-dimensional scalar theory with fields
valued in a generic non-commutative algebra. We have emphasized that a closed RG equation exists
for the functional generator of connected amputated graphs with sphere topology and all possible
configurations of external legs, which represent multi-traces of the field sources. The crucial
difference between the ordinary and the planar Wilson-Polchinski equations lies in the quantum (loop)
term: it is proportional to second order field derivatives in the ordinary equation and to first
order field derivatives in the planar one. Thus the planar RG equation for a $d$-dimensional system
has the form of a Hamilton-Jacobi equation for a non-relativistic system in $d+1$ dimensions, whose time
is identified with the renormalization scale.

This result is of course reminiscent of the holographic conjecture. However a precise relation ---
if it exists at all --- between our $d$-dimensional RG equation and the Hamilton-Jacobi equation 
for $d+1$-dimensional 
supergravity of \cite{Verlinde} remains unclear to us. In particular, it should be stressed that Wilson RG,
even in Polchinski's differential form, describes an invariance property of the Feynman integral and hence concerns
the effective action which is a functional of the elementary fields while
the object that enters the holographic conjecture is a vacuum energy, functional of the sources of composite 
operators.

The planar Wilson-Polchinski equation should be a useful tool to investigate the renormalization 
properties of field theories over non-commutative space. The 
equation describes the regime when external momenta are large in comparison with $1/\sqrt{|\Theta|}$, with 
$\Theta$ the non-commutative space-time tensor, and thus it does not concern the 
$|\Theta|\to 0$ limit. As first explained in \cite{Seiberg}, this limit
is made subtle by the fact that non-planar Green functions contains negative powers of $\Theta$. 
This does not necessarily imply the presence of singularities of the renormalized theory in the limit 
$|\Theta|\to 0$. Indeed the limit of small non-commutativity should be studied by renormalizing the 
theory at an energy scale is much smaller than $1/\sqrt{|\Theta|}$. In this framework deformations
corresponding to the small $|\Theta|$ expansion of the star product should represent irrelevant terms. 
From this point of view the UV/IR paradox just originates from a not-allowed limit exchange.

As a last application, we have shown that in the case of $0$-dimensional hermitian matrix models the 
large $N$ RG equation can be recast in a very simple form: it is equivalent to the {\it free} 
Hamilton-Jacobi 
equation for a system of particles in 1 dimension. This reformulation has lead to a closed expression
for the large $N$ effective action of the model, which is equivalent to the free energy of one-matrix 
model coupled to an external matrix source. As far as we know, the expression in
Eqs. (\ref{finaleffectiveaction}), (\ref{algebraiceq}) for the free energy of 
this model with {\it arbitrary} potential represents a new result.

\section*{Acknowledgments}

We thank M. Porrati for helpful discussions. This work is supported in part by the European Commission's 
Human Potential
programme under contract HPRN-CT-2000-00131 ``The quantum structure of spacetime and the geometric 
nature of fundamental interactions'', to which the authors are associated through the Frascati National 
Laboratory, and by Ministero dell'Universit\`a e della Ricerca Scientifica e Tecnologica.

\appendix

\section{Non-Planar Non-Commutative Graphs}

To understand how one can approach  a field theory over a 
non-commutative space-time
using our RG equation in the framework of perturbation
theory it can be useful to present the first steps of the
construction.

The starting point is the introduction of a cut-off propagator that
we choose according to:
\be
\D_\La (p^{2}) = \! \int_{\La_0^{-2}}^{\La^{-2}}\! dx\, {\rm e}^{-x\,p^2}
\ee
Here
$\La_{0}$ is the UV cut-off and $\La$ is the running cut-off that is identified with the RG parameter.
Let ${1/\sqrt{|\Theta|}}$ be the
non-commutative mass scale. 
We will consider the evolution equation in the regime
\be 
{1\o\sqrt{|\Theta|}}<\La\ll\La_0
\ee
The effective action $H_\La[\varphi]$ is computed as a series expansion of the 
coupling $\la$ starting from the initial condition
\be 
H_{\La_0} [\varphi]={\la\o4}\!\int\! \prod_{i = 1}^4 dp_{\sss 4,\, i}\, \d
\bigl(\sum_{j = 1}^4 p_{\sss 4,\, j}\bigr)\, \langle \varphi (p_{\sss 4,\, 1})\cdots
\varphi (p_{\sss 4,\, 4 })\rangle \,{\rm e}^{i\,\Phi(p_{\sss 4,\,1} \ldots\, p_{\sss 4,\, 4} )}
\ee
where we have defined the Moyal factor
\be
\Phi(p_{\sss k,\,1}\ldots p_{\sss k,\, k}) \equiv \sum_{i=2}^{k-1}\Bigl(\sum_{j=1}^{i-1}\,
p_{\sss k,\, j}\Bigr) \Theta\, p_{\sss k,\, i}
\ee 
The terms in the effective action of Eq.~(\ref{scfunctional}) proportional to the amplitudes
$h^{\{\d_{k,2}\}}_{\La} (\pk)$ and $h^{\{\d_{k,6}\}}_{\La}$ are determined at the first 
perturbative order to be:
\bea 
\la\!\int\! \prod_{i=1}^2
\, dp_{\sss 2,\, i}\, \d \bigl(\sum_{j=1}^2p_{\sss 2,\,j}\bigr)\Biggl[\int\! {dp\o (2\pi)^d} 
\int_{\La_0^{-2}}^{\La^{-2}}\! dx\,  {\rm e}^{-x\,p^2}+C\Biggr]
\langle
\varphi (p_{\sss 2,\, 1})\, \varphi (p_{\sss 2,\, 2 })
\rangle \nn 
+ \, {\la^2\o 2}\!\int\!\prod_{i=1}^6 \, dp_{\sss 6,\, i}\int\! dp\, \d 
\bigl(\sum_{j=1}^3 p_{\sss 6,\, j} + p \bigr) \d \bigl(\sum_{l=4}^6 p_{\sss 6,\,l} - 
p\bigr)\int_{\La_0^{-2}}^{\La^{-2}}\! dx\,
{\rm e}^{-x\,p^2}
\nn \times\, {\rm e}^{i\,\Phi (p_{\sss 6,\,1} \ldots\, p_{\sss 6,\, 6} )}\, \langle
\varphi (p_{\sss 6,\, 1})\cdots \varphi (p_{\sss 6,\, 6})\rangle
\eea 
These contributions correspond, in the zero dimensional case, to the terms linear in $t$ 
in Eq. (\ref{hsol}).
The constant $C$ in the first term of equation above appears due to the 
UV divergence of the corresponding diagram; it must be
determined using suitable normalization conditions at some fixed 
$\La_R >1/\sqrt{|\Theta|}$. For example, if we require $h^{\{\d_{k,2}\}}_{\La}$  to
vanish at  $p_{\sss 2,\,1} = 0$ at the normalization point $\La_R$, the term in brackets 
$\int\! {dp\o (2\pi)^d} \int_{\La_0^{-2}}^{\La^{-2}}\! dx\, {\rm e}^{-x\,p^2}+C$ becomes
$(4\pi)^{-{d\o2}}\,(1-{d\o2})^{-1}\,\bigl(\La^{d-2}-\La_R^{d-2}\bigl)$, which is UV safe.

With a further step we reach the terms corresponding  to the $t^2$ 
contributions in (\ref{hsol}). At this order the first non-planar term 
in the perturbative expansion of the effective action appears: 
\be
\int\!\!\!\prod_{i_2,j_2=1}^2
\!\!\! dp^{\sss (i_2)}_{{\scriptscriptstyle 2,\, j_2}}\,\d\bigl(\!\!\sum_{ i_2,j_2=1}^2\, 
p^{\sss (i_2)}_{{\scriptscriptstyle 2,\, j_2}}\bigr)\,
h^{\{2\d_{k,2}\}}_{\La}(\pk)\,
\langle\varphi (p^{\sss (1)}_{\sss 2,\, 1})
\varphi (p^{\sss (1)}_{\sss 2,\,2 })
\rangle\, \langle\varphi (p^{\sss (2)}_{\sss 2,\, 1})
\varphi (p^{\sss (2)}_{\sss 2,\,2 })
\rangle
\ee
The non-planar amplitude $h^{\{2\d_{k,2}\}}_{\La} (\pk)$ receives contributions fron the two
following diagrams (with inward oriented  momenta):

\begin{figure}[h]
\centerline{
\begin{picture}(60,30)(0,0)
\put(0,6){\line(1,0){20}}
\put(0,10){\makebox(0,0){\tiny $p_1$}}
\put(55,10){\makebox(0,0){\tiny $p_2$}}
\put(20,10){\makebox(0,0){\tiny $p_3$}}
\put(35,10){\makebox(0,0){\tiny $p_4$}}
\put(28,6){\circle{32}}
\put(35,6){\line(1,0){20}}
\end{picture}
\hskip 1cm
\begin{picture}(60,30)(0,0)
\put(12,-11){\line(1,0){32}}
\put(12,-6){\makebox(0,0){\tiny $p_1$}}
\put(44,-6){\makebox(0,0){\tiny $p_2$}}
\put(20,7){\makebox(0,0){\tiny $p_3$}}
\put(38,7){\makebox(0,0){\tiny $p_4$}}
\put(28,6){\circle{32}}
\put(18,12){\line(1,1){10}}\put(28,22){\line(1,-1){10}}
\end{picture}
}
\end{figure}

\noindent Let us consider only the first diagram whose contribution to 
$h^{\{2\d_{k,2}\}}_{\La}$ is:
\begin{eqnarray} 
&&\!\!\!\!\!\!\!\!\!\!h^{\{2\d_{k,2}\}}_{\La,1}\! = {\la^2\over 4}\! \int\! {dk\o (2\pi)^d}\, 
{\rm e}^{i\,2\,k\,\Theta(p_1+p_2)}\!\int_{\La_0^{-2}}^{\La^{-2}}\! dx\, dy\,
{\rm e}^{-\bigl[x\,\bigl(k+{p_2+p_4\over 2}\bigr)^2 + \,y\,\bigl(k-{p_2+p_4\over 2}\bigr)^2\bigr]}\\
&&\!\!\!\!\!= {\la^2\o 4\,(4\pi)^{{d\o2}}}\!\int_{\La_0^{-2}}^{\La^{-2}}\!\! {dx\, dy\o(x+y)^{d\o2}}\,
{\rm e}^{-{1\o(x+y)}\bigl[xy\,(p_2+p_4)^2 +(\Theta(p_1+p_2))^2 
+i (x-y)(p_2+p_4)\,\Theta(p_1+p_2)\bigr]}\nonumber
\end{eqnarray}

In the non-exceptional regime, in which
\be 
{1\o\sqrt{|\Theta|}}<\La < |p_1+p_2|\sim |p_2+p_4|
\ee 
one can safely take the
(UV) limit $\La_{0}\rightarrow\infty$ and, after a variable rescaling, one obtains
\bea 
\!\!\!\!\!\!\!\!\!\!\!\!\!\!\!\!h^{\{2\d_{k,2}\}}_{\La,1}
= {\la^2\o 4\,(4\pi)^{{d\o2}}|\Theta|^{{d\o2}-2}}
\int_0^{1\o |\Theta|\La^2}\!{dx\,dy\o(x+y)^{d\o2}}
\nn\,\,\,\,\times\, 
{\rm e}^{-{1\o(x+y)}\bigl[xy\,|\Theta|(p_2+p_4)^2 + {\(\Theta(p_1+p_2)\)^2\o|\Theta|} +
i(x-y)(p_2+p_4)\,\Theta(p_1+p_2)\bigr]}
\eea
Hence asymptotically, that is when ${1\o\sqrt{|\Theta|}}<\La \ll |p_1+p_2|\sim |p_2+p_4|$, we find
\bea 
h^{\{2\d_{k,2}\}}_{\La,1}\simeq {\la^2\o 4\,(4\pi)^{{d\o2}}|\Theta|^{{d\o2}-2}}
 \int_0^{1\o |\Theta|\La^2}\!{dx\,dy\o (x+y)^{d\o2}}\,{\rm e}^{-{1\o(x+y)}
{\(\Theta(p_1+p_2)\)^2\o|\Theta|}}
\nn\qquad\quad\,\, \simeq {\la^2\o 4\,(4\pi)^{d\o2}\({\La^2\o2}\)^{2-{d\o2}}} 
\int_1^2 \!dz\, z^{{d\o2}-2}\,
{\rm e}^{-\bigl({1\over 2}\Theta(p_1+p_2)\bigr)^2\La^2\,z}\nn 
\qquad\quad\,\, \simeq {\la^2\o 4\,(4\pi)^{d\o2}\({\La^2\o2}\)^{2-{d\o2}}}\,
{\rm e}^{-\({1\over 2}\Theta(p_1+p_2)\)^2\La^2}
\label{ris}
\eea
This example confirms that when external momenta are non-exceptional non planar 
amplitudes are exponentially suppressed as noted in the Introduction.

One could wonder how it is possible that non-planar terms contribute 
to the evolution equation of planar ones --- the observation of which fact is the
main result of this paper. Our example clarifies also this point. 
Indeed the hole amplitude that we have studied contributes
to the evolution equation of the two-loop self energy by a term in 
which two legs belonging to the same hole are connected by
$\dot\D$. However --- and this is the crucial remark --- this situation does not
belong to the non-exceptional regime since $p_3+p_4=-p_1-p_2=0$. In this case it is apparent 
that the exponential damping factor in (\ref{ris}) disappears.

\section{Differential Equations for Matrix Models}

Let $F(\{g_n\})$ be the vacuum energy of $0$-dimensional matrix model
with an infinite number of single-trace interactions
\be
{\rm e}^{N^2\,F(\{g_n\})}\equiv \int d\Phi\, {\rm e}^{N\,S_1(\Phi;\{g_n\})}
\ee 
where
\be
S_1(\Phi;\{g_n\}) = \sum_{n=1}^\infty g_n\,\Tr\, \Phi^n  
\ee
The scale invariance of the matrix functional measure implies that 
\be
e^{-N^2\,F}\int d\Phi\,\Tr\, \partial_\Phi \bigl(\Phi\, {\rm e}^{N\,S_1}\bigr) = 0
\ee
which translates in the following first order differential
equation for $F(\{g_n\})$
\be
1 + \sum_{n=1}^{\infty} n\,g_n\,\partial_{g_{n}} F = 0
\label{lineareq}
\ee
A similar first order linear differential equation follows from translation invariance of the
functional measure.
In an analogous way we can derive a non-linear differential equation for
$F(\{g_n\})$ starting from
\be
e^{-N^2\,F}\int d\Phi\,\Tr\, \partial^2_\Phi {\rm e}^{N\, S_1} = 
N\,\langle \Tr\, \partial^2_\Phi\, S_1 \rangle \, +\, N^2\,\langle \Tr\, (\partial_\Phi\,S_1)^2\rangle = 0
\ee
Using the relations
\bea
\langle \Tr\,(\partial_\Phi\, S_1)^2 \rangle = N\sum_{n,\,m = 1}^\infty \,n\, m \,g_n\, g_m \,
\partial_{g_{n+m-2}} F\nn
\langle \Tr \,\partial^2_\Phi\, S_1 \rangle = N^2 \sum_{n=1}^\infty\sum_{l=0}^{n-2}n\,g_n\,\Bigl(
\partial_{g_l} F \,\partial_{g_{n-2-l}} F + {1\over N^2}\,\partial_{g_l} \partial_{g_{n-2-l}} F \Bigr)
\eea
(and reorganizing the sums) we find, in the large $N$ limit
\be
\sum_{n,\,m = 0}^\infty\!\!\! (n+m+2)\, g_{n+m+2}
\,\partial_{g_n}F \, \partial_{g_m}F + \sum_{n = 0}^\infty \sum_{l = 1}^{n+1}\! l\, 
(n-l+2)\, g_{l}\, g_{n-l+2}\,\partial_{g_n}F = 0
\label{miaolieq}
\ee
where we define $\partial_{g_0} F \equiv 1$.

Eq. (\ref{miaolieq}) allows to recursively determine $F(\{g_n\})$ as a perturbative series in 
$g_2^{-1}$.
In order to make the recursive computation simple it is convenient to introduce the following 
functional
\be
{\hat F}_t(\{g_n\})\equiv -{1\over 2}\,\log\,t + F\bigl(\{g_n-{1\over 2t}\d_{n,2}\})
\label{fhat}
\ee 
which is essentially the functional considered in \cite{MiaoLi}. More precisely, the functional
of \cite{MiaoLi} coincides with ${\hat F}_t(\{g_n\})$ after the couplings $g_n$ with $n\le 2$ have been
set to zero. Note however that a simple differential equation can only be written for 
${\hat F}_t(\{g_n\})$ as a functional of all the couplings $g_n$ with $n\ge 1$. The definition 
Eq.~(\ref{fhat}) and the two differential equations (\ref{lineareq}) and (\ref{fhat}) yield
\be
\partial_t\,{\hat F_t}\!=\! {1\over 2}\!\!\sum_{\,\,n,\,m = 0}^\infty\!\!\!\! (n+m+2)\, g_{n+m+2}
\,\partial_{g_n}{\hat F} \, \partial_{g_m}{\hat F} + {1\over 2} \sum_{n = 0}^\infty 
\sum_{l = 1}^{n+1}\!l\, (n-l+2)\, g_{l}\, g_{n-l+2}\,\partial_{g_n}{\hat F}
\label{miaolieqtwo}
\ee
which is the equation derived in \cite{MiaoLi}. Eq.~(\ref{miaolieqtwo}) and the initial condition
\be
{\hat F}_t(\{g_n\})|_{t=0} = 0
\ee
determine ${\hat F}_t(\{g_n\})$ as a power series in $t$. The first terms of the series are
\be
{\hat F_t} = \Bigl(g_2 + {1\over 2} g_1^2 \Bigr) t + \Bigl( g_2^2 + g_1^2 g_2 + {3\over 2} g_1 g_3 + {3\over 2} g_6 
+ {5\over 2} g_1 g_5 + g_1^2 g_4 \Bigr) t^2 + O(t^3)
\ee

\end{document}